\documentclass[prd,a4paper,showpacs]{revtex4}
\usepackage{epsfig}
\usepackage{color}
\usepackage{amssymb}
\usepackage{amsfonts}
\usepackage{graphicx}
\usepackage{keyval,graphicx}
\usepackage{textcomp,wasysym}

\begin{document}
\title{Study of $\eta_c$ and $\eta_c^\prime$ decays into vector meson pairs }
\author{Qian Wang$^1$, Xiao-Hai Liu$^1$, Qiang Zhao$^{1,2}$}

\affiliation{1) Institute of High Energy Physics, Chinese Academy of
Sciences, Beijing 100049, P.R. China \\
2) Theoretical Physics Center for Science Facilities, CAS, Beijing
100049, China}

\begin{abstract}

The processes of $\eta_c (\eta_c')\to VV$ are supposed to be
suppressed by the helicity selection rule (HSR) but found to be
rather important decay modes in experiment for $\eta_c$. We try to
distinguish the short-distance transitions via the singly
disconnected Okubo-Zweig-Iizuka (SOZI) processes and the doubly
disconnected (DOZI) processes in $\eta_c (\eta_c')\to VV$. It shows
that the SOZI processes can be related to the $\eta_c(\eta_c')$
wavefunctions at the origin. Therefore, a relation of the decay
branching ratio fraction between these two decay channels can be
established. Such a relation is similar to that for $J/\psi$ and
$\psi^\prime\to VP$, where the so-called ``$\rho\pi$ puzzle" has
been existed for a while. We also show that the intermediate charmed
meson loop transitions provide an evading mechanism for the DOZI
processes. This contribution would turn out to be more important in
$\eta_c^\prime\to VV$. As a consequence, it may produce significant
deviations from the SOZI-dominant scenario. Future experimental
measurement of the $\eta_c^\prime\to VV$ by BESIII should be able to
clarify the DOZI-evading mechanisms.

\end{abstract}

\date{\today}

\maketitle

\section{Introduction}

Among the charmonium states below $D\bar{D}$ threshold, there is
little information on the $\eta_c'$ decays from experiment.  Even
its mass and width still have large uncertainties as listed in the
Particle Data Group \cite{Amsler:2008zzb}. From a theoretical view
point, this state contains rich information on the dynamics in the
interplay of perturbative and non-perturbative QCD, for which there
are still a lot of questions unanswered.

Our motivation of studying $\eta_c$ and $\eta_c'$ simultaneously is
driven by two puzzling but interesting questions. Firstly, it has
been observed that $\eta_c\to VV$, where $V$ stands for light vector
meson, is one of the most important decay channels for $\eta_c$ with
branching ratios at the order of $10^{-3}$ to $10^{-2}$. However,
these decay channels as well as $\eta_c'\to VV$ are supposed to be
highly suppressed by the so-called helicity selection rule
(HSR)~\cite{Brodsky:1981kj,Chernyak:1981zz,Chernyak:1983ej}. Such an
observation of the HSR violation indicates the importance of QCD
higher twist contributions or presence of a non-pQCD mechanism that
violates the HSR. The contradiction between the data and the HSR
expectations based on the perturbative QCD (pQCD) has drawn some
attention, and various attempts have been made to try to understand
the underlying
dynamics~\cite{Benayoun:1990ey,Anselmino:1990vs,Zhou:2005fc,Anselmino:1993yg,Zhao:2006cx,Braaten:2000cm,Santorelli:2007xg,Gong:2008ue,Sun:2010qx}.
In Refs.~\cite{Liu:2009vv,Liu:2010um}, we studied the effects of
charmed hadron loops as a source of long distance contribution which
violates the HSR. The results indeed suggest that contributions from
intermediate charmed hadron loops are significant and can be taken
as an evading mechanism of the HSR. In $\eta_c(\eta_c')\to VV$ it is
natural to expect that the charmed hadron loops would contribute and
might provide an evading mechanism here. Even for higher energy
processes such as $\eta_b\to J/\psi J/\psi$, it was also found that
charmed meson loops would enhance the decay rate
significantly~\cite{Santorelli:2007xg}. In Ref.~\cite{Zhang:2008ab},
possible contributions from charmed meson loops to $e^+ e^-\to
J/\psi\eta_c$ at $W=10.56$ GeV were also investigated.

The second reason that we are interested in $\eta_c(\eta_c')$
exclusive decays is that they are related with the long-standing
so-called ``$\rho\pi$ puzzle" in $J/\psi (\psi')\to VP$. The decays
of $J/\psi$ and $\psi'$   into light hadrons are supposed to be via
the valence $c\bar{c}$ annihilations into three gluons in pQCD to
the leading order at a typical distance of $1/m_c$. In the heavy
quark limit, i.e. $m_c$ is infinitely large, the decay amplitude
will be proportional to the wave function at the origin $\psi(0)$
($\psi'(0)$). As a result the following relation is expected for the
inclusive final states of light hadrons $h$,
\begin{eqnarray}
R_{\psi\psi'} \equiv \frac{BR(\psi'\to h)}{BR(J/\psi \to h)} =
\frac{BR(\psi'\to e^+e^-)}{BR(J/\psi \to e^+e^-)}
 \simeq
\left|\frac{\psi'(0)}{\psi(0)} \right|^2
\frac{\Gamma_{tot}^{\psi}}{\Gamma_{tot}^{\psi'}} \simeq 12\%,
\label{12-percent}
\end{eqnarray}
where the phase space factors are taken into account by
$\Gamma_{tot}$ in the heavy quark limit. For the inclusive decay and
many exclusive decays, the data exhibit consistencies with this
relation rather well, which is called ``$12\%$ rule". However, some
exclusive decay modes, such as $\rho\pi$ and $K^*\bar{K}+c.c.$, are
found largely violating the ``$12\%$ rule", which is known as the
``$\rho\pi$ puzzle" problem.

Many theoretical efforts have been made in order to understand the
origin of such a significant deviation from the ``12\% rule" in
$J/\psi (\psi')\to \rho\pi$ and $K^*\bar{K}+c.c.$ In recent
works~\cite{Zhao:2006gw,Li:2007ky}, we show that the interferences
between the strong and EM decay amplitudes in both $J/\psi
(\psi')\to VP$ are essential for understanding the ``$\rho\pi$
puzzle". Similar ideas had been proposed in the
literature~\cite{Suzuki:2001fs,Seiden:1988rr}. However, if this is
the case, one has to clarify and even quantify the mechanism which
suppressed the strong transition amplitude in $\psi'\to VP$ and then
makes it to be compatible with the EM amplitudes in some channels. A
numerical study of the overall decay channels for $J/\psi (\psi')\to
VP$ indeed suggests such a phenomenon~\cite{Li:2007ky}. In
Refs.~\cite{Li:2007ky,zhao-QCD2010}, it was shown that the
intermediate charmed meson loops, which serves as a long-distance
mechanism for both OZI rule and HSR evasions, can significantly
suppress the $\psi'\to VP$ strong transition amplitudes. Such a
mechanism will largely alter the branching ratios for $J/\psi
(\psi')\to VP$, and makes the relation in Eq.~(\ref{12-percent})
unreliable.

In fact, the intermediate charmed meson loops could be much more
general than we would expect in charmonium energy region. Further
studies of the effects from intermediate meson loops in other
processes have been reported in
Refs.~\cite{Zhang:2009kr,Liu:2009vv,Liu:2009dr,Guo:2010zk,Guo:2010ak,Liu:2010um}.

The coherent study of the $\eta_c$ and $\eta_c'\to VV$ will be
useful for clarifying the role played by the intermediate meson
loops and EM transition amplitudes in $J/\psi (\psi')\to VP$. Since
$\eta_c$ and $\eta_c'$ are just the spin 0 partners of $J/\psi$ and
$\psi'$ respectively, and they may possess the same spatial wave
functions in the heavy quark limit, we would expect a similar
relation as Eq.(\ref{12-percent}) to hold between $\eta_c$ and
$\eta_c'$, namely,
\begin{eqnarray}
R_{\eta_c\eta_c'} \equiv \frac{BR(\eta_c'\to h)}{BR(\eta_c \to h)} =
\frac{BR(\eta_c'\to \gamma\gamma)}{BR(\eta_c \to \gamma\gamma)}
 \simeq
\left|\frac{\eta_c'(0)}{\eta_c(0)} \right|^2
\frac{\Gamma_{tot}^{\eta_c}}{\Gamma_{tot}^{\eta_c'}} \simeq 0.52
\sim 1.56,
\end{eqnarray}
where we have taken $\eta_c'(0)/\eta_c(0)=\psi'(0)/\psi(0)=0.64$
\cite{Hou:1982kh}, and the range of the ratio is displayed
considering the larger uncertainties of the $\eta_c'$ total
width~\cite{Amsler:2008zzb}. Alternatively, we can express this
relation as
\begin{equation}
\bar{R}_{\eta_c\eta_c'}\equiv \frac{\Gamma_{\eta_c'\to
h}}{\Gamma_{\eta_c \to h}}\simeq \left|\frac{\eta_c'(0)}{\eta_c(0)}
\right|^2\simeq 0.41 \ ,\label{retac}
\end{equation}
which avoids the uncertainties with the total width of
$\eta_c^\prime$.

In exclusive decays of $\eta_c (\eta_c')\to VV$, as discussed
earlier the transition amplitudes should be suppressed by the HSR.
However, since the mass of the charm quark is not heavy enough and
due to the non-vanishing light quark masses, the HSR violation
should occur via both singly disconnected Okubo-Zweig-Iizuka (SOZI)
transitions and doubly disconnected (DOZI) processes. The latter can
be related to long-distance hadronic transitions via intermediate
charmed meson loops~\cite{Zhang:2009kr}. For the SOZI transitions,
the relation of Eq.~(\ref{retac}) will be respected due to their
short-distance feature, while the long-distance charmed meson loops
will not necessarily respect it. In parallel with the observations
in $J/\psi(\psi^\prime)\to VP$, if the charmed meson loops play a
more significant role in $\eta_c^\prime\to VV$ than in $\eta_c\to
VV$, they may violate the relation of Eq.~(\ref{retac}) and lead to
observable phenomena similar to the ``$\rho\pi$ puzzle" in these two
decay channels.

Our focus in this work is to investigate the role played by the
intermediate charmed meson loops in $\eta_c(\eta_c^\prime)\to VV$.
Also, we mention in advance that due to lack of experimental
information to constrain the relative strength between SOZI and loop
transition amplitudes, some of the results have to be qualitative.
But they can be examined by the forthcoming BESIII high-statistics
experiment.

As follows, in Sec. II, we discuss the parametrization scheme for
$\eta_c (\eta_c^\prime)\to VV$ and constraints from the available
experimental data. The intermediate charmed meson loops are
described by an effective Lagrangian approach in Sec. III, and
numerical results are presented. A brief summary is given in Sec.
IV.

\section{Parametrization for $\eta_c (\eta_c^\prime)\to VV$}

An obvious advantage for the exclusive decays of $\eta_c
(\eta_c^\prime)\to VV$ is that these transitions, similar to
$J/\psi(\psi^\prime)\to VP$, have only one unique Lorentz structure
for the $VVP$ couplings. As stressed a number of times before, this
will allow a parametrization of the effective coupling constant
contributed by different mechanisms. This will help a lot especially
at this moment when the experimental data for $\eta_c^\prime$ are
not available.

In Ref.~\cite{Zhao:2006cx}, a parametrization scheme is proposed to
distinguish the SOZI and DOZI processes by the gluon counting rule.
Due to the above mentioned property with the $VVP$ coupling, one can
factorize out the DOZI evading amplitude, of which the ratio to the
SOZI amplitude will give an estimate of the order of magnitude of
the contributions from the DOZI mechanisms.

Following Ref.~\cite{Zhao:2006cx}, the transition amplitudes for
$\eta_c\to VV$ can be expressed as
\begin{eqnarray}
\label{isospin-1} \langle \phi\phi | \hat{V}_{gg}| \eta_c\rangle
&=& g_0^2 R^2 (1+r)   \nonumber\\
\langle \omega\omega | \hat{V}_{gg}| \eta_c\rangle
&=& g_0^2(1+2r)  \nonumber\\
\langle \omega\phi | \hat{V}_{gg}| \eta_c\rangle &=& g_0^2 r R
\sqrt{2}
\nonumber\\
\langle K^{*+}K^{*-}|  \hat{V}_{gg}| \eta_c\rangle
&=& g_0^2 R \nonumber\\
\langle \rho^+\rho^- | \hat{V}_{gg}| \eta_c\rangle &=& g_0^2  \ .
\end{eqnarray}
where $\hat{V}_{gg}$ is the $\eta_c\to gg\to (q\bar{q})(q\bar{q})$
potential, and parameter $g_0$ denotes the coupling strength of the
SOZI transitions.  Parameter $r$ is the ratio of the DOZI transition
over the SOZI transition. It should be pointed out that the
additional gluon exchange in DOZI is not necessarily perturbative.
Due to contributions from effective mesonic degrees of freedom,
namely, intermediate meson exchanges, the DOZI transitions can be
evaded by rather soft gluon exchanges of which the contributions can
be parameterized by $r$. We also introduce the SU(3) flavor breaking
parameter $R$, of which its deviation from unity reflects the change
of couplings due to the mass difference between $u/d$ and $s$. The
amplitudes for other charge combinations of $K^*\bar{K^*}$ and
$\rho\rho$ are implicated.

A commonly used form factor is adopted in the calculation of the
partial decay widths:
\begin{equation}
{\cal F}^2({\bf p})=p^{2l}\exp(-{\bf p}^2/8\beta^2) \ ,
\end{equation}
where ${\bf p}$ and $l$ are the three momentum and relative angular
momentum of the final-state mesons, respectively, in the $\eta_c$
rest frame. We adopt $\beta=0.5$ GeV, which is the same as in
Refs.~\cite{close-amsler,close-kirk,close-zhao-f0,zhao-chi-c}. Such
a form factor will largely account for the size effects from the
spatial wavefunctions of the initial and final state mesons.

The present experimental situation should be clarified. In 2005,
BESII Collaboration measured the exclusive decay branching ratios of
$\eta_c\to VV$~\cite{Ablikim:2005yi}, where they find significant
differences from the DM2 results~\cite{Bisello:1990re} in $\eta_c\to
\rho\rho$. In fact, the PDG averaged value for $\eta_c\to \rho\rho$
branching ratio is strongly affected by this discrepancy. As a
comparison of the BESII results and the PDG averaged values, we list
in Table ~\ref{tab-1} the fitting results of BESII data (Fit-II in
Ref.~\cite{Zhao:2006cx}) and PDG2008~\cite{Amsler:2008zzb}. The
fitted parameters are listed in Table~\ref{tab-2}.


\begin{table}[ht]
\begin{tabular}{c|c|c||c|c}
\hline BR $(\times 10^{-3})$ & \ \ \
BESII~\protect\cite{Ablikim:2005yi} \ \ \
& \ \ \ Fit-BES \ \ \ & PDG2008~\protect\cite{Amsler:2008zzb} & \ \ \ Fit-PDG \ \ \ \\[1ex]
\hline
$\rho\rho$ & $12.5\pm 3.7\pm 5.1$ & $10.7$ & $20.0\pm 7.0$ & 11.80 \\[1ex]
$K^*\bar{K^*}$ & $10.4\pm 2.6\pm 4.3$ & $13.6$  & $9.2\pm 3.4$ & 11.76 \\[1ex]
$\phi\phi$ & $2.5\pm 0.5\pm 0.9$ & $2.16$ & $2.7\pm 0.9$ & 2.23 \\[1ex]
$\omega\omega$ &  $<6.3$ & \ \ \ $1.67$  & $< 3.1$ & 4.53 \\[1ex]
$\omega\phi$ & $<1.3 $ & \ \ \ $0.33$  & $<1.7$ & 0.02 \\[1ex]
\hline
\end{tabular}
\caption{ The branching ratios for $\eta_c\to VV$. The data are from
BES. Fit-BES is obtained by fitting the BESII
data~\protect\cite{Ablikim:2005yi} for $\eta_c\to \phi\phi$,
$K^*\bar{K^*}$ and $\rho\rho$, while Fit-PDG are obtained by fitting
the PDG2008 data~\protect\cite{Amsler:2008zzb}. } \label{tab-1}
\end{table}

\begin{table}[ht]
\begin{tabular}{c|c|c}
\hline
Parameters & \ \ \  Fit-BES \ \ \ & \ \ \ Fit-PDG \ \ \ \\[1ex]
\hline $r$ & \ \ \   $-0.16\pm 0.15$ \ \ \ &  \ \ \  $0.04\pm 0.16$  \ \ \    \\[1ex]\hline
$R$ & $1.02\pm 0.23$   & $0.91\pm 0.16$  \\[1ex]\hline
$g_0$ & $0.35\pm 0.04$ & $0.36\pm 0.04$ \\[1ex]\hline
$\chi^2$ & $0.5$  & 2.4  \\[1ex]\hline
\end{tabular}
\caption{ The parameters determined in Fit-BES and Fit-PDG. }
\label{tab-2}
\end{table}

It is interesting to read that the two fitting results are rather
consistent with each other in $\eta_c\to \rho\rho$, $K^*\bar{K^*}$
and $\phi\phi$. Also, the fitted parameters agree well in these two
fits except that the  $\chi^2$ has a much larger value in Fit-PDG.
The fitted branching ratio for $\rho\rho$ has significant
discrepancies with the PDG averaged value. Differences between these
two sets of data lead to a relatively large uncertainty with
parameter $r$, while parameters $g_0$ and $R$ are rather stable. It
suggests that the DOZI-evading mechanism will only account for an
order of $r^2\simeq 0\sim 0.1$ in $\eta_c\to VV$, and the SOZI
process should be dominant. This observation is within our
expectation and consistent with what we find in $J/\psi
(\psi^\prime)\to VP$, where the DOZI-evading mechanism in $J/\psi\to
VP$ is negligibly small~\cite{Li:2007ky}.

The fitting results also show that the present experimental
uncertainties are rather large, and the DOZI-evading mechanism
cannot be well constrained by the data. It should be pointed out
that the relation of Eq.~(\ref{retac}) may help estimate the SOZI
contributions in $\eta_c^\prime\to VV$ given the dominance of the
SOZI processes. However, due to lack of data for $\eta_c^\prime$, it
is not possible to estimate the DOZI evading contributions in
$\eta_c^\prime\to VV$ based on the parametrization scheme. In order
to proceed, we assume that the DOZI-evading mechanisms are via the
intermediate charmed meson loops, for which a quantitative study of
the effects can be quantified. By taking the upper limit of the DOZI
contributions in $\eta_c\to VV$, we can then estimate the possible
impact of the DOZI-evading mechanisms in $\eta_c(\eta_c^\prime)\to
VV$. In particular, we would like to examine whether the relation of
Eq.~(\ref{retac}) still holds or not with the presence of the DOZI
contributions.

\section{DOZI-evading mechanisms via intermediate charmed meson loops}

\subsection{Formulation}

\begin{figure}[tb]
\begin{center}
\begin{tabular}{cc}
 \hspace{-5.5cm}\includegraphics[scale=0.5]{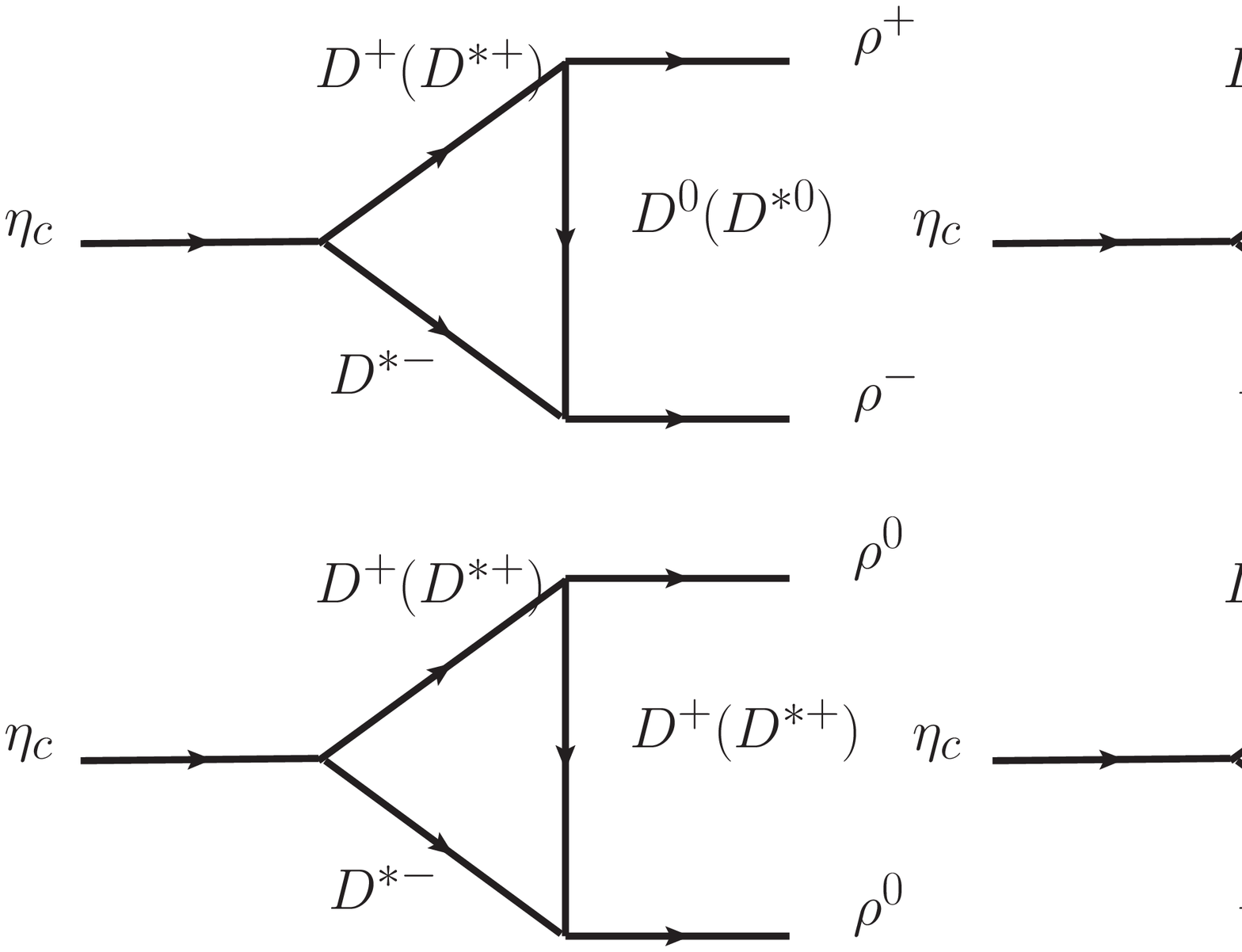}
\end{tabular}
\caption{Feynman diagrams for $\eta_c\to \rho\rho$ via intermediate
charmed meson loops.}\label{fig-rho}
\end{center}
\end{figure}

\begin{figure}[tb]
\begin{center}
\begin{tabular}{cc}
 \hspace{-5.5cm}\includegraphics[scale=0.5]{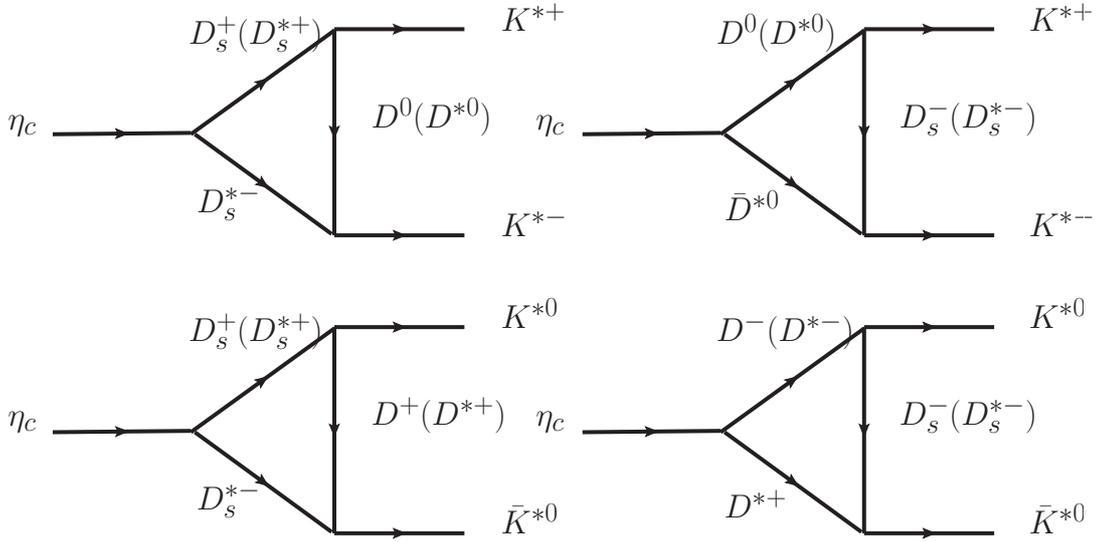}
\end{tabular}
\caption{Feynman diagrams for $\eta_c\to K^*\bar{K}^*$ via
intermediate charmed meson loops.}\label{fig-k}
\end{center}
\end{figure}

\begin{figure}[tb]
\begin{center}
\hspace{-6cm}
\includegraphics[scale=0.5]{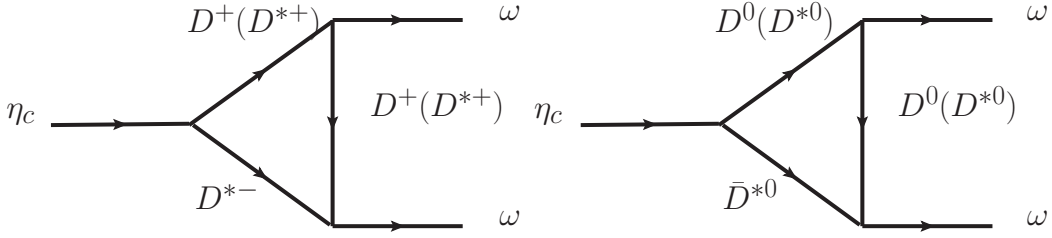}
 \caption{Feynman diagrams for $\eta_c\to
\omega\omega$ via intermediate charmed meson
loops.}\label{fig-omega}
\end{center}
\end{figure}

\begin{figure}[tb]
\begin{center}
\hspace{-2cm}
\includegraphics[scale=0.5]{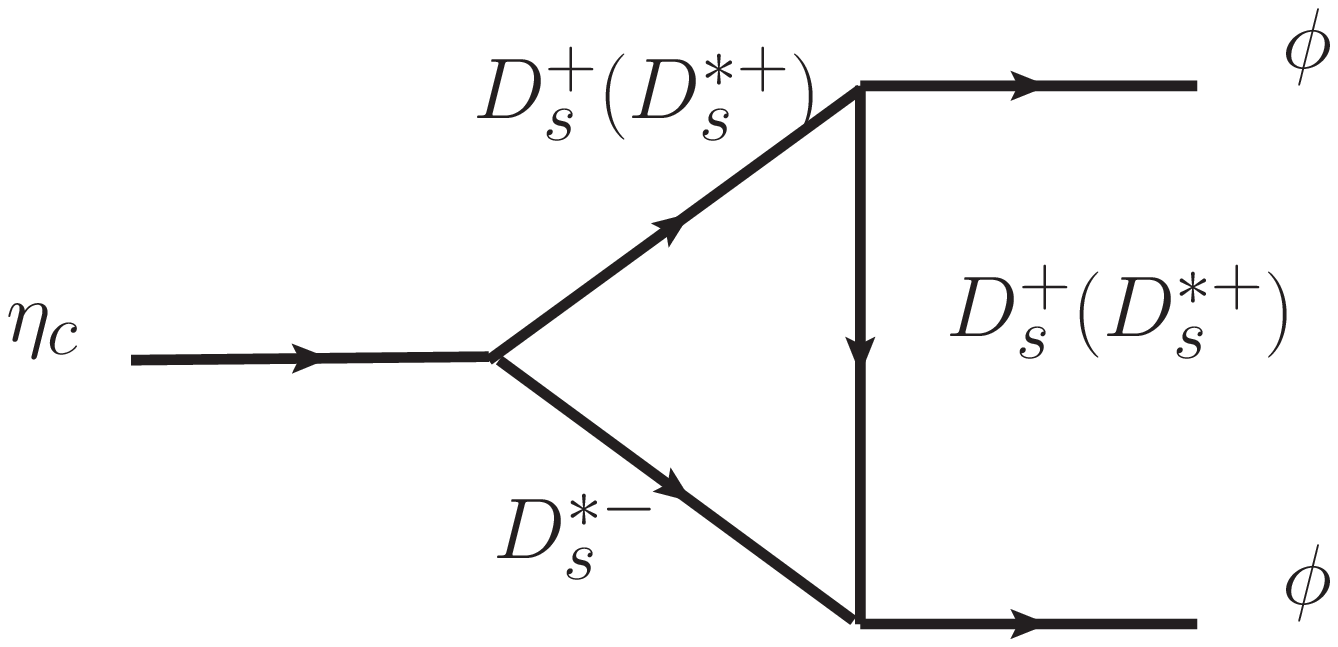}
\caption{Feynman diagrams for $\eta_c\to \phi\phi$ via intermediate
charmed meson loops.}\label{fig-phi}
\end{center}
\end{figure}

We will use an effective Lagrangian approach to estimate the
transition amplitudes. In Figs.~\ref{fig-rho}, \ref{fig-k},
\ref{fig-omega} and \ref{fig-phi}, the Feynman diagrams of $\eta_c$
decaying to $\rho\rho$, $K^*\bar{K}^*$, $\omega\omega$, and
$\phi\phi$ via the intermediate charmed meson loops are presented
respectively. The relevant Lagrangians based on heavy quark symmetry
which describes the coupling between $S$-wave charmonium and the
charmed mesons reads~\cite{Colangelo:2003sa,Casalbuoni:1996pg},

\begin{equation}
\mathcal{L}_2=i g_2 Tr[R_{c\bar{c}} \bar{H}_{2i}\gamma^\mu
{\stackrel{\leftrightarrow}{\partial}}_\mu \bar{H}_{1i}] + H.c.,
\end{equation}
where the $S$-wave charmonium states are expressed as
\begin{equation}
R_{c\bar{c}}=\left( \frac{1+ \rlap{/}{v} }{2} \right)\left(\psi^\mu
\gamma_\mu -\eta_c \gamma_5 \right )\left( \frac{1- \rlap{/}{v} }{2}
\right).
\end{equation}
And the charmed and anti-charmed meson triplet are
\begin{eqnarray}
H_{1i}&=&\left( \frac{1+ \rlap{/}{v} }{2} \right)[
\mathcal{D}_i^{*\mu}
\gamma_\mu -\mathcal{D}_i\gamma_5], \\
H_{2i}&=& [\bar{\mathcal{D}}_i^{*\mu} \gamma_\mu
-\bar{\mathcal{D}}_i\gamma_5]\left( \frac{1- \rlap{/}{v} }{2}
\right),
\end{eqnarray}
where $\mathcal{D}$ and $\mathcal{D}^*$ are pseudoscalar
($(D^{0},D^{+},D_s^{+})$) and vector charmed mesons
($(D^{*0},D^{*+},D_s^{*+})$), respectively. The Lagrangian
describing the interactions between light mesons and charmed mesons
reads
\begin{eqnarray}
 {\cal L} &=&\nonumber
 -ig_{\rho\pi\pi}\Big(\rho^+_\mu\pi^0{\stackrel{\leftrightarrow}{\partial}}{\!^\mu}\pi^-
 +\rho^-_\mu\pi^+{\stackrel{\leftrightarrow}{\partial}}{\!^\mu}\pi^0+\rho^0_\mu\pi^-{\stackrel{\leftrightarrow}{\partial}}{\!^\mu}\pi^+\Big)
  \\\nonumber
 &-& ig_{\mathcal{D}^*\mathcal{D}\mathcal{P}}(\mathcal{D}^i\partial^\mu \mathcal{P}_{ij}
 \mathcal{D}_\mu^{*j\dagger}-\mathcal{D}_\mu^{*i}\partial^\mu \mathcal{P}_{ij}\mathcal{D}^{j\dagger})
 +{1\over 2}g_{\mathcal{D}^*\mathcal{D}^*\mathcal{P}}
 \epsilon_{\mu\nu\alpha\beta}\,\mathcal{D}_i^{*\mu}\partial^\nu \mathcal{P}^{ij}
 {\stackrel{\leftrightarrow}{\partial}}{\!^\alpha} \mathcal{D}^{*\beta\dagger}_j
 \\\nonumber
 &-& ig_{\mathcal{D}\mathcal{D}\mathcal{V}} \mathcal{D}_i^\dagger {\stackrel{\leftrightarrow}{\partial}}{\!_\mu} \mathcal{D}^j(V^\mu)^i_j
 -2f_{\mathcal{D}^*\mathcal{D}\mathcal{V}} \epsilon_{\mu\nu\alpha\beta}
 (\partial^\mu \mathcal{V}^\nu)^i_j
 (\mathcal{D}_i^\dagger{\stackrel{\leftrightarrow}{\partial}}{\!^\alpha} \mathcal{D}^{*\beta j}-\mathcal{D}_i^{*\beta\dagger}{\stackrel{\leftrightarrow}{\partial}}{\!^\alpha} D^j)
 \\
 &+& ig_{\mathcal{D}^*\mathcal{D}^*\mathcal{V}} \mathcal{D}^{*\nu\dagger}_i {\stackrel{\leftrightarrow}{\partial}}{\!_\mu} \mathcal{D}^{*j}_\nu(\mathcal{V}^\mu)^i_j
 +4if_{\mathcal{D}^*\mathcal{D}^*\mathcal{V}} \mathcal{D}^{*\dagger}_{i\mu}(\partial^\mu \mathcal{V}^\nu-\partial^\nu
 \mathcal{V}^\mu)^i_j \mathcal{D}^{*j}_\nu,
 \label{eq:LDDV}
 \end{eqnarray}
with the convention $\epsilon^{0123}=+1$, where $\mathcal{P}$ and
$\mathcal{V}_\mu$ denote $3\times 3$ matrices for the pseudoscalar
octet and vector nonet mesons respectively~\cite{Cheng:2004ru},
\begin{eqnarray}
 \mathcal{P} &=& \left(\matrix{{\pi^0\over\sqrt{2}}+{\eta\over\sqrt{6}} & \pi^+ & K^+ \cr
 \pi^- & -{\pi^0\over\sqrt{2}}+{\eta\over\sqrt{6}} & K^0  \cr
 K^- & \bar{ K^0} & -\sqrt{2\over 3}\eta }\right), \ \
\mathcal{V} =
\left(\matrix{{\rho^0\over\sqrt{2}}+{\omega\over\sqrt{2}} & \rho^+ &
K^{*+} \cr
 \rho^- & -{\rho^0\over\sqrt{2}}+{\omega\over\sqrt{2}} & K^{*0}  \cr
 K^{*-} & \bar{ K^{*0}} & \phi }\right).
 \end{eqnarray}

The following kinematic conventions are adopted,  $\eta_c(p)\to
\mathcal{D}^{(*)}(p_1)\bar{\mathcal{D}}^{(*)}(p_3)[\mathcal{D}^{(*)}(p_2)]\to
V(k)V(q)$, where $\mathcal{D}^{(*)}$ in the square bracket denotes
the exchanged charmed meson in the triangle diagrams. There are four
types of transition amplitudes corresponding to different kinds of
charmed meson exchanges in the triangle diagrams. We take the
amplitudes of $\eta_c\to \phi\phi$ as an example,
\begin{eqnarray}
\mathcal{M}_{\mathcal{D}\mathcal{D}^*[\mathcal{D}]}&=&\int\frac{d^4p_1}{(2\pi)^4}[-2g_{\eta_c\mathcal{D}\mathcal{D}^*}g_{\mathcal{D}\mathcal{D}\mathcal{V}}f_{\mathcal{D}\mathcal{D}^*\mathcal{V}}\epsilon_{\mu\nu\alpha\beta}(p_1+p_2)\cdot\epsilon_kq^\mu\epsilon_q^\nu
\nonumber\\
&\times&(p_2-p_3)^\alpha(p_1-p_3)_\lambda
(-g^{\lambda\beta}+\frac{p_3^\lambda p_3^\beta}{m_3^2})]\frac{1}{a_1
a_2 a_3}\mathcal{F}(p_i^2),\\\nonumber
\mathcal{M}_{\mathcal{D}\mathcal{D}^*[\mathcal{D}^*]}&=&\int\frac{d^4p_1}{(2\pi)^4}[-8g_{\eta_c\mathcal{D}\mathcal{D}^*}f_{\mathcal{D}\mathcal{D}^*\mathcal{V}}f_{\mathcal{D}^*\mathcal{D}^*\mathcal{V}}\epsilon_{\mu\nu\alpha\beta}k^\mu\epsilon_k^\nu(p_1+p_2)^\alpha
\\\nonumber&\times&(-g^{\beta\lambda}+\frac{p_2^\beta p_2^\lambda}{m_2^2})
(p_1-p_3)_\delta (-g^{\delta\theta}+\frac{p_3^\delta
p_3^\theta}{m_3^2})(\epsilon_{q\lambda}q_\theta-q_\lambda\epsilon_{q\theta})
\\\nonumber
&+&2g_{\eta_c\mathcal{D}\mathcal{D}^*}f_{\mathcal{D}\mathcal{D}^*\mathcal{V}}g_{\mathcal{D}^*\mathcal{D}^*\mathcal{V}}\epsilon_{\mu\nu\alpha\beta}k^\mu\epsilon_k^\nu(p_1+p_2)^\alpha(p_2-p_3)\cdot\epsilon_q
\\&\times&(p_1-p_3)_\delta(-g^{\beta\lambda}+\frac{p_2^\beta
p_2^\lambda}{m_2^2})(-g_\lambda^{\delta}+\frac{p_{3\lambda}
p_3^\delta}{m_3^2})]\frac{1}{a_1 a_2
a_3}\mathcal{F}(p_i^2),\\\nonumber
\mathcal{M}_{\mathcal{D}^*\mathcal{D}^*[\mathcal{D}]}&=&\int\frac{d^4p_1}{(2\pi)^4}[-4g_{\eta_c\mathcal{D}^*\mathcal{D}^*}f_{\mathcal{D}^*\mathcal{D}^*\mathcal{V}}^2\epsilon_{\mu\nu\alpha\beta}
\epsilon_{\rho\sigma\lambda\iota}\epsilon_{\eta\theta\xi\omega}p^\nu
p_1^\mu  (p_1+p_2)^\lambda \\&\times&k^\rho \epsilon_k^\sigma q^\eta
\epsilon_q^\theta (p_2-p_3)^\xi (-g^{\beta\iota}+\frac{p_1^\beta
p_1^\iota}{m_1^2})(-g^{\alpha\omega}+\frac{p_3^\alpha
p_3^\omega}{m_3^2})]\frac{1}{a_1 a_2 a_3}\mathcal{F}(p_i^2),\\
\mathcal{M}_{\mathcal{D}^*\mathcal{D}^*[\mathcal{D}^*]}&=&\int\frac{d^4p_1}{(2\pi)^4}[\mathcal{A}_1+\mathcal{A}_2+\mathcal{A}_3+\mathcal{A}_4]\frac{1}{a_1
a_2 a_3}\mathcal{F}(p_i^2),
\\
\nonumber
 \mathcal{A}_1&=&g_{\eta_c\mathcal{D}^*\mathcal{D}^*}g_{\mathcal{D}^*\mathcal{D}^*\mathcal{V}}^2\epsilon_{\mu\nu\alpha\beta}p^\nu p_1^\mu(p_1+p_2)\cdot\epsilon_k
 (p_2-p_3)\cdot \epsilon_q \\\nonumber
 &\times&(-g^{\beta\lambda}+\frac{p_1^\beta
 p_1^\lambda}{m_1^2})(-g_{\lambda\delta}+\frac{p_{2\lambda}
 p_{2\delta}}{m_2^2})(-g^{\delta\alpha}+\frac{p_3^\delta
 p_3^\alpha}{m_3^2})\\
 \nonumber
 \mathcal{A}_2&=&-4g_{\eta_c\mathcal{D}^*\mathcal{D}^*}g_{\mathcal{D}^*\mathcal{D}^*\mathcal{V}}f_{\mathcal{D}^*\mathcal{D}^*\mathcal{V}}\epsilon_{\mu\nu\alpha\beta}p^\nu
 p_1^\mu (p_1+p_2)\cdot \epsilon_k\\
 \nonumber&\times& (-g^{\beta\lambda}+\frac{p_1^\beta
 p_1^\lambda}{m_1^2})(-g_{\lambda\rho}+\frac{p_{2\lambda}
 p_{2\rho}}{m_2^2})(-g^{\alpha}_\delta+\frac{p_{3\delta}
 p_3^\alpha}{m_3^2})(\epsilon_q^\rho q^\delta-q^\rho
 \epsilon_q^\delta)\\
 \nonumber
 \mathcal{A}_3&=&-4g_{\eta_c\mathcal{D}^*\mathcal{D}^*}g_{\mathcal{D}^*\mathcal{D}^*\mathcal{V}}f_{\mathcal{D}^*\mathcal{D}^*\mathcal{V}}\epsilon_{\mu\nu\alpha\beta}p^\nu
 p_1^\mu (-p_3+p_2)\cdot \epsilon_q \\
 \nonumber&\times&(-g^{\beta}_\lambda+\frac{p_1^\beta
 p_{1\lambda}}{m_1^2})(-g_{\sigma\rho}+\frac{p_{2\sigma}
 p_{2\rho}}{m_2^2})(-g^{\rho\alpha}+\frac{p_3^\rho
 p_3^\alpha}{m_3^2})(k^\sigma \epsilon_k^\lambda-\epsilon_k^\sigma k^\lambda)\\
 \nonumber
 \mathcal{A}_4&=&16g_{\eta_c
 \mathcal{D}^*\mathcal{D}^*}f_{\mathcal{D}^*\mathcal{D}^*\mathcal{V}}^2\epsilon_{\mu\nu\alpha\beta}p^\nu p_1^\mu (-g^{\beta}_\lambda+\frac{p_1^\beta
 p_{1\lambda}}{m_1^2})(-g^{\alpha}_\rho+\frac{p_2^\alpha
 p_{2\rho}}{m_2^2})\\\nonumber&\times&(-g_{\theta\delta}+\frac{p_{3\theta}
 p_{3\delta}}{m_3^2})(q^\rho \epsilon_k^\lambda k^\theta \epsilon_q^\delta +
\epsilon_q^\rho k^\lambda q^\theta \epsilon_k^\delta
 -\epsilon_q^\rho \epsilon_k^\lambda q^\theta k^\delta - q^\rho k^\lambda \epsilon_k^\theta \epsilon_q^\delta)
\end{eqnarray}
where $a_1\equiv p_1^2-m_1^2, \ a_2\equiv p_2^2-m_2^2$, and
$a_3\equiv p_3^2-m_3^2$. The amplitudes of other processes have
similar forms as the above. We omit them for the sake of brevity.

Since the couplings in the effective Lagrangians are local ones,
ultra-violate divergence in the loop integrals is inevitable. We
introduce a form factor $\mathcal{F}(p_i^2)$ phenomenologically to
take into account the non-local effects and cut off the divergence
in the loop integrals, i.e.
\begin{equation}
\mathcal{F}(p_i^2)=\prod\limits_i\left(
\frac{\Lambda_i^2-m_i^2}{\Lambda_i^2-p_i^2} \right ),
\end{equation}
where $m_i$ and $p_i$ are the mass and four momentum of the
corresponding exchanged particle, and the cut-off energy is chosen
as $\Lambda_i=m_i+\alpha\Lambda_{QCD}$ with $\Lambda_{QCD}=0.22$ GeV
\cite{Cheng:2004ru,Liu:2009vv,Liu:2010um}. The value of parameter
$\alpha$ is commonly expected to be of order of unity.

\subsection{Numerical results for DOZI-evading contributions}

Before proceeding to the numerical results, we first determine some
of the parameters taken in this approach. In the chiral and heavy
quark limit, the following relations can be
obtained~\cite{Cheng:2004ru,Casalbuoni:1996pg},
\begin{eqnarray}
g_{\mathcal{D}\mathcal{D}\mathcal{V}}=g_{\mathcal{D}^*\mathcal{D}^*\mathcal{V}}=\frac{\beta
g_\mathcal{V}}{\sqrt{2}}, \
f_{\mathcal{D}\mathcal{D}^*\mathcal{V}}=\frac{f_{\mathcal{D}^*\mathcal{D}^*\mathcal{V}}}{m_{\mathcal{D}^*}}=\frac{\lambda
g_\mathcal{V}}{\sqrt{2}}, \
 g_\mathcal{V}=\frac{m_\rho}{f_\pi}, \\
g_{\eta_c \mathcal{D}\mathcal{D}^*}=g_{\eta_c \mathcal{D}^*
\mathcal{D}^*}\sqrt{\frac{m_\mathcal{D}}{m_{\mathcal{D}^*}}}m_{\eta_c}
=2g_2\sqrt{m_{\eta_c}m_{\mathcal{D}}m_{\mathcal{D}^*}},
\end{eqnarray}
where $\beta$ and $\lambda$ are commonly taken as $\beta=0.9,
\lambda=0.56$ GeV$^{-1}$, while $f_\pi$ is the pion decay constant.

In principle, the coupling $g_2$ should be computed by
nonperturbative methods. If we simply estimate the coupling $g_2$
with the vector meson dominance (VMD) argument, it will give
$g_2=\sqrt{m_\psi}/(2m_\mathcal{D} f_\psi)$, where $m_\psi$ and
$f_\psi=405$ MeV being the mass and decay constant of $J/\psi$
\cite{Colangelo:2003sa}. This relation gives $g_{\eta_c \mathcal{D}
\mathcal{D}^*}=7.68$, which is a commonly adopted value in the
literature. In order to determine $g_{\eta_c^\prime
\mathcal{D}^{(*)} \mathcal{D}^{(*)} }$, we similarly relate it to
the coupling of $g_{\psi^\prime
\mathcal{D}\mathcal{D}}=9.05$~\cite{Zhang:2009gy}, which is
determined by the experimental data for $e^+ e^-\to
D\bar{D}$~\cite{Pakhlova:2008zza}. In Ref.~\cite{Li:2007xr}, similar
values for $g_{\eta_c\mathcal{D}^{(*)} \mathcal{D}^{(*)}}$ and
$g_{\eta_c^\prime \mathcal{D}^{(*)} \mathcal{D}^{(*)}}$ were
adopted.

The determination of the form factor parameter $\alpha$ will depend
on the relative strengths and interferences between the SOZI and
DOZI-evading amplitudes. As discussed in the previous Section, due
to lack of sufficient experimental information about the
$\eta_c^\prime\to VV$, we can only estimate the upper limit of the
DOZI-evading contributions based on the data for $\eta_c\to VV$.
Interestingly, we find that for the same value of the form factor
parameter $\alpha$, the ratio $\Gamma^{loop}_{\eta_c^\prime\to
VV}/\Gamma^{loop}_{\eta_c\to VV}$ is rather insensitive to $\alpha$
of a broad range. As mentioned before, the advantage of taking this
ratio is that the uncertainties of the $\eta_c^\prime$ total width
can be avoided. In Fig.~\ref{fig-ratio-width}, the ratio
$\Gamma^{loop}_{\eta_c^\prime\to VV}/\Gamma^{loop}_{\eta_c\to VV}$
in terms of a common range of $\alpha$ is displayed. The
dot-dot-dashed line (independent of $\alpha$) denotes the ratio
given by Eq.~(\ref{retac}), which turns to be much smaller than the
ratio between the charmed meson loops.

The following points can be learned:

\begin{figure}[tb]
\begin{center}
\includegraphics[scale=0.5]{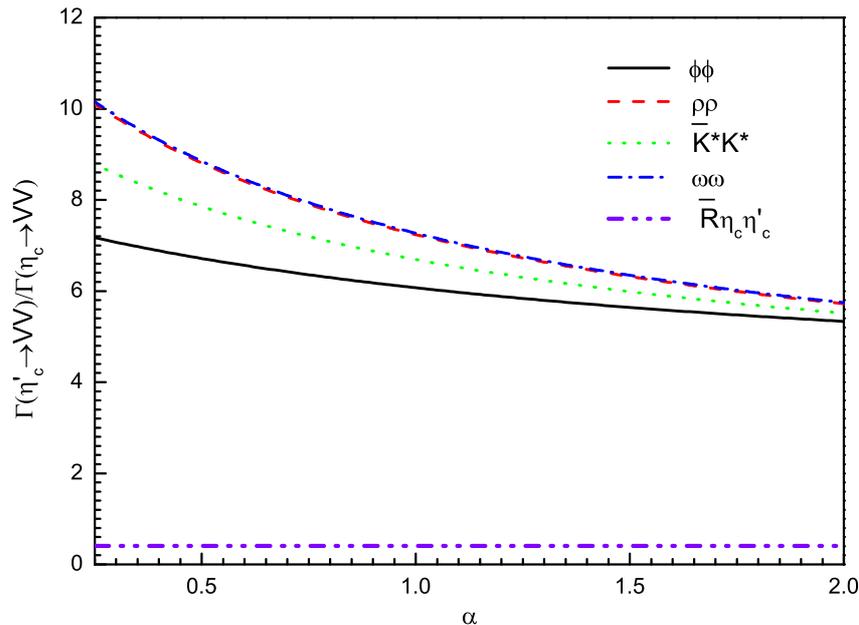}
 \caption{The partial width fractions $\Gamma^{loop}_{\eta_c^\prime \to VV}/\Gamma^{loop}_{\eta_c \to
 VV}$ in terms of the parameter $\alpha$. The ratio of
 $\bar{R}_{\eta_c\eta_c'}$, which is independent of $\alpha$, is also displayed by the dot-dot-dashed
 line.
}\label{fig-ratio-width}
\end{center}
\end{figure}

\begin{figure}[tb]
\begin{center}
\hspace{-4cm}
\includegraphics[scale=0.5]{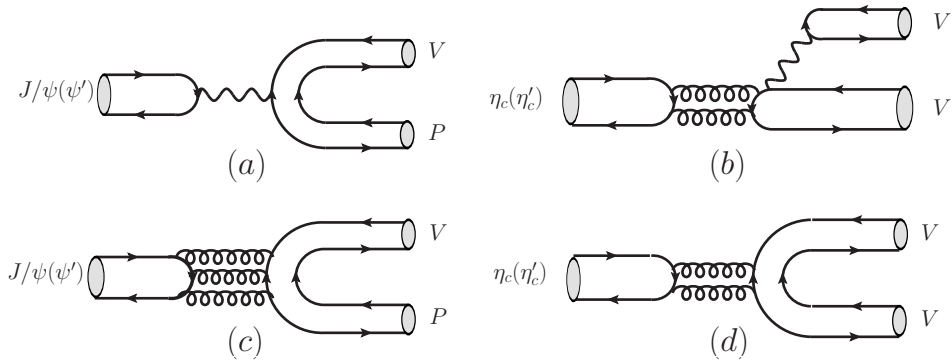}
 \caption{Schematic diagrams for the EM and strong
decays of $J/\psi (\psi^\prime)\to VP$ and $\eta_c(\eta_c^\prime)\to
VV$. }\label{fig-power}
\end{center}
\end{figure}

\begin{center}
\begin{table}
\begin{tabular}{cccc}
  \hline\hline
  $\Gamma^{loop} (\times 10^{-2})$ MeV & $\eta_c\to VV$ & $\eta_c^\prime\to VV$ \\
 \hline
  $\phi\phi$ & $0.48\sim 0.96$ & $3.10 \sim 6.05$  \\
  \hline
  $\rho\rho$ & $3.50 \sim 6.87$ & $28.04 \sim 53.70$ \\
  \hline
  $K^*\bar{K}^*$ & $3.12 \sim 6.20$ & $23.10 \sim 44.40$ \\
  \hline
  $\omega\omega$ & $1.15\sim 2.28$ & $9.33\sim 17.90$  \\
  \hline\hline
\end{tabular}
\caption{The DOZI-evading contributions to the partial widths of
$\eta_c(\eta_c^\prime)\to VV$ estimated by the charmed meson loops.
The parameter $\alpha$ varies between $0.69 \sim 0.78$, which
produces about $10\%$ of the experimental data for $\eta_c\to VV$ as
estimated by the parametrization scheme. } \label{dozi-evading}
\end{table}
\end{center}

i) In case that the decays of $\eta_c\ (\eta_c^\prime)\to h$ to be
dominated by the short-distance transitions, i.e. SOZI process, one
expects that the relation of Eq.~(\ref{retac}) would be respected,
i.e.
\begin{equation}
\Gamma_{\eta_c^\prime\to VV}\simeq
\left|\frac{\eta_c^\prime(0)}{\eta_c(0)}\right|^2\times\Gamma_{\eta_c\to
VV}\simeq 0.41 \times\Gamma_{\eta_c\to VV} \ , \label{ratio-width}
\end{equation}
where the small DOZI-evading contributions in $\eta_c\to VV$ are
neglected.

ii) In case that the DOZI-evading contributions in $\eta_c\to VV$
amount to about $10\%$ of the SOZI as shown by the fitting results
of the previous section, it shows that the intermediate charmed
meson loops will be sizeable in $\eta_c^\prime\to VV$.  In
Table~\ref{dozi-evading}, we list the DOZI-evading contributions to
the partial widths in comparison with the SOZI-dominant predictions.
These two mechanisms seem to be compatible in $\eta_c^\prime$, which
may easily violate the relation of Eq.~(\ref{retac}). Such an effect
can be directly examined by experimental data for the
$\eta_c^\prime\to VV$. Such a possibility that the DOZI-evading
contributions are sizeable in $\eta_c^\prime\to VV$ exhibits
similarities as in $\psi^\prime\to VP$, where the charmed meson
loops play an important role of causing the deviations from the
``12\% rule".

iii) It should be mentioned again that the decays
$\eta_c(\eta_c^\prime)\to VV$ are strong-interaction-dominant
processes, which are different from the decays $J/\psi (\psi^\prime)
\to VP$ where the EM interaction will also play an important role,
especially in $\psi^\prime\to VP$~\cite{Zhao:2006gw,Li:2007ky}. We
illustrate this point in Fig.~\ref{fig-power}. By these diagrams, a
naive power counting indicates that
\begin{eqnarray}
\frac{T_{em}}{T_{str}} &\sim& \frac{\alpha_e}{\alpha_s^2}\ \
\mbox{for}\ \  J/\psi (\psi') \to VP, \nonumber\\
\frac{T_{em}}{T_{str}} &\sim& \alpha_e\ \ \mbox{for}\ \ \eta_c
(\eta_c')\to VV,
\end{eqnarray}
where ``$T_{em}$" and ``$T_{str}$" denote the leading EM and strong
transition amplitudes, respectively. It implies that the EM
contribution in $\eta_c (\eta_c')\to VV$ will be less important than
that in $J/\psi (\psi') \to VP$.

An interesting expectation that distinguishes the role played by the
EM transitions in $\eta_c(\eta_c^\prime)\to VV$ and
$J/\psi(\psi^\prime)\to VP$ is that the long-distance corrections
from the charmed meson loops should have the same relative phases in
different exclusive decay channels. This means that the possible
deviations from Eq.~(\ref{ratio-width}) caused by the charmed meson
loops will either enhance or lower the overall $\eta_c^\prime\to VV$
branching ratios. In contrast, in $J/\psi(\psi^\prime)\to VP$,
interferences due to the EM transitions may have different phases in
different channels, e.g. the branching ratios of $\psi^\prime\to
K^{*0}\bar{K^0}+c.c.$ is much enhanced in comparison with
$\psi^\prime\to K^{*+} K^- +c.c.$~\cite{Amsler:2008zzb}. Therefore,
observation of systematic deviations from Eq.~(\ref{ratio-width})
would be a strong evidence of contributions from intermediate
charmed meson loops.

\section{summary}

In this paper, we have investigated the processes of $\eta_c
(\eta_c')\to VV$. These decay modes are supposed to be highly
suppressed by the HSR but found to be rather important according to
the available data. The intermediate charmed meson loop transitions,
which are correlated with the DOZI-evading, are introduced to
provide a mechanism for the HSR violation, and the results indicate
that the effect of these loops may be significant, especially in
$\eta_c^\prime\to VV$.

Although we are still lack of experimental information to constrain
the model parameters in these transitions, we show that a
parametrization scheme based on the unique $VVP$ coupling structure
will allow us to parameterize out the SOZI and DOZI-evading
transitions in $\eta_c\to VV$. By assuming that the DOZI-evading
processes can be recognized by the intermediate charmed meson loops
as a soft long-distance transition, we find that the SOZI
contributions in $\eta_c\to VV$ can be well constrained by the
experimental data, and an upper limit for the DOZI-evading
contributions can be set. This allows us to proceed and estimate the
DOZI-evading contributions in $\eta_c^\prime\to VV$. It shows that
the intermediate charmed meson loops may produce measurable effects
in $\eta_c^\prime\to VV$ and cause sizeable deviations from the
relation predicted by the dominance of the SOZI transitions.

This situation is similar to the exclusive processes of $J/\psi
(\psi')\to VP$. There, the intermediate charmed meson loops provide
a natural explanation for the so-called ``$\rho\pi$ puzzle". In this
sense, the study of $\eta_c(\eta_c^\prime)\to VV$ would provide
additional evidence for the intermediate charmed meson loops in
charmonium decays, and can be examined quantitatively by the future
BESIII experiment~\cite{bes-iii}.

We should also mention that our conclusions are based on the
hypothesis that the flavor components of $\eta_c$ and $\eta_c'$ are
dominated by $c\bar{c}$. Thus, the connection of their spatial
wavefunctions with those of $J/\psi$ and $\psi'$ will make sense. If
$\eta_c$ or $\eta_c'$ possesses some other internal
structures~\cite{Feldmann:2000hs,Anselmino:1993yg,Anselmino:1990vs},
the relation between their branching ratios will be affected to some
extent, which however, is not our focus in this work.

\section{acknowledgements}

The authors thank X.-Y. Shen for useful information about the BESII
analysis of $\eta_c\to VV$. This work is supported, in part, by the
National Natural Science Foundation of China (Grants No. 10675131
and 10491306), Chinese Academy of Sciences (KJCX3-SYW-N2), and
Ministry of Science and Technology of China (2009CB825200).


\begin{thebibliography}{99}

\bibitem{Amsler:2008zzb}
  C.~Amsler {\it et al.}  [Particle Data Group],
  Phys.\ Lett.\  B {\bf 667}, 1 (2008).

\bibitem{Brodsky:1981kj}
  S.~J.~Brodsky and G.~P.~Lepage,
  Phys.\ Rev.\  D {\bf 24}, 2848 (1981).

\bibitem{Chernyak:1981zz}
  V.~L.~Chernyak and A.~R.~Zhitnitsky,
  Nucl.\ Phys.\  B {\bf 201}, 492 (1982)
  [Erratum-ibid.\  B {\bf 214}, 547 (1983)].

\bibitem{Chernyak:1983ej}
  V.~L.~Chernyak and A.~R.~Zhitnitsky,
  Phys.\ Rept.\  {\bf 112}, 173 (1984).




\bibitem{Benayoun:1990ey}
  M.~Benayoun, V.~L.~Chernyak and I.~R.~Zhitnitsky,
  Nucl.\ Phys.\  B {\bf 348} (1991) 327.




\bibitem{Anselmino:1990vs}
  M.~Anselmino, F.~Murgia and F.~Caruso,
  Phys.\ Rev.\  D {\bf 42}, 3218 (1990).




\bibitem{Anselmino:1993yg}
  M.~Anselmino, M.~Genovese and D.~E.~Kharzeev,
  Phys.\ Rev.\  D {\bf 50}, 595 (1994)
  [arXiv:hep-ph/9310344].

\bibitem{Zhao:2006cx}
  Q.~Zhao,
  Phys.\ Lett.\  B {\bf 636}, 197 (2006)
  [arXiv:hep-ph/0602216].

\bibitem{Zhou:2005fc}
  H.~Q.~Zhou, R.~G.~Ping and B.~S.~Zou,
  Phys.\ Rev.\  D {\bf 71}, 114002 (2005).

\bibitem{Braaten:2000cm}
  E.~Braaten, S.~Fleming and A.~K.~Leibovich,
  Phys.\ Rev.\  D {\bf 63}, 094006 (2001)
  [arXiv:hep-ph/0008091].


\bibitem{Santorelli:2007xg}
  P.~Santorelli,
  Phys.\ Rev.\  D {\bf 77}, 074012 (2008)
  [arXiv:hep-ph/0703232].






\bibitem{Gong:2008ue}
  B.~Gong, Y.~Jia and J.~X.~Wang,
  Phys.\ Lett.\  B {\bf 670}, 350 (2009)
  [arXiv:0808.1034 [hep-ph]].



\bibitem{Sun:2010qx}
  P.~Sun, G.~Hao and C.~F.~Qiao,
  arXiv:1005.5535 [hep-ph].

\bibitem{Liu:2009vv}
  X.~H.~Liu and Q.~Zhao,
  Phys.\ Rev.\  D {\bf 81}, 014017 (2010)
  [arXiv:0912.1508 [hep-ph]].

\bibitem{Liu:2010um}
  X.~H.~Liu and Q.~Zhao,
  arXiv:1004.0496 [hep-ph].

\bibitem{Zhang:2008ab}
  Y.~J.~Zhang, Q.~Zhao and C.~F.~Qiao,
  Phys.\ Rev.\  D {\bf 78}, 054014 (2008)
  [arXiv:0806.3140 [hep-ph]].

\bibitem{Zhao:2006gw}
  Q.~Zhao, G.~Li and C.~H.~Chang,
  Phys.\ Lett.\  B {\bf 645}, 173 (2007)
  [arXiv:hep-ph/0610223].


\bibitem{Li:2007ky}
  G.~Li, Q.~Zhao and C.~H.~Chang,
  J.\ Phys.\ G {\bf 35}, 055002 (2008)
  [arXiv:hep-ph/0701020]; Q. Zhao {\it et al.}, Chinese Phys. {\bf
  C 34}, 299 (2010).

\bibitem{Suzuki:2001fs}
  M.~Suzuki,
  Phys.\ Rev.\  D {\bf 63}, 054021 (2001).

\bibitem{Seiden:1988rr}
  A.~Seiden, H.~F.~W.~Sadrozinski and H.~E.~Haber,
  Phys.\ Rev.\  D {\bf 38}, 824 (1988).

\bibitem{zhao-QCD2010} Q. Zhao, Proceeding of QCD2010, Montpellier,
France, 2010.


\bibitem{Zhang:2009kr}
  Y.~J.~Zhang, G.~Li and Q.~Zhao,
  Phys.\ Rev.\ Lett.\  {\bf 102}, 172001 (2009)
  [arXiv:0902.1300 [hep-ph]].


\bibitem{Liu:2009dr}
  X.~Liu, B.~Zhang and X.~Q.~Li,
  Phys.\ Lett.\  B {\bf 675}, 441 (2009)
  [arXiv:0902.0480 [hep-ph]].



\bibitem{Guo:2010zk}
  F.~K.~Guo, C.~Hanhart, G.~Li, U.~G.~Meissner and Q.~Zhao,
  Phys.\ Rev.\  D {\bf 82}, 034025 (2010)
  [arXiv:1002.2712 [hep-ph]].


\bibitem{Guo:2010ak}
  F.~K.~Guo, C.~Hanhart, G.~Li, U.~G.~Meissner and Q.~Zhao,
  arXiv:1008.3632 [hep-ph].

\bibitem{Hou:1982kh}
  W.~S.~Hou and A.~Soni,
  Phys.\ Rev.\ Lett.\  {\bf 50}, 569 (1983).
%
\bibitem{zhao-chi-c} Q. Zhao, Phys. Rev. D {\bf 72}, 074001
(2005); hep-ph/0508086.
%
\bibitem{close-amsler} C. Amsler and F.E. Close,
Phys.\ Lett.\ B {\bf 353}, 385 (1995); Phys. Rev. {\bf D53}, 295
(1996).
%
\bibitem{close-kirk} F.E. Close and A. Kirk,
Phys.\ Lett.\ B {\bf 483}, 345 (2000).
%
\bibitem{close-zhao-f0} F.E. Close and Q. Zhao,
Phys. Rev. D {\bf 71}. 094022 (2005) [arXiv:hep-ph/0504043].

\bibitem{Ablikim:2005yi}
  M.~Ablikim {\it et al.}  [BES Collaboration],
  Phys.\ Rev.\  D {\bf 72}, 072005 (2005)
  [arXiv:hep-ex/0507100].


\bibitem{Bisello:1990re}
  D.~Bisello {\it et al.}  [DM2 collaboration],
  Nucl.\ Phys.\  B {\bf 350}, 1 (1991).

\bibitem{Colangelo:2003sa}
  P.~Colangelo, F.~De Fazio and T.~N.~Pham,
  Phys.\ Rev.\  D {\bf 69}, 054023 (2004)
  [arXiv:hep-ph/0310084].



\bibitem{Casalbuoni:1996pg}
  R.~Casalbuoni, A.~Deandrea, N.~Di Bartolomeo, R.~Gatto, F.~Feruglio and G.~Nardulli,
  Phys.\ Rept.\  {\bf 281}, 145 (1997)
  [arXiv:hep-ph/9605342].






\bibitem{Cheng:2004ru}
  H.~Y.~Cheng, C.~K.~Chua and A.~Soni,
  Phys.\ Rev.\  D {\bf 71}, 014030 (2005)
  [arXiv:hep-ph/0409317].










\bibitem{Zhang:2009gy}
  Y.~J.~Zhang and Q.~Zhao,
  Phys.\ Rev.\  D {\bf 81}, 034011 (2010)
  [arXiv:0911.5651 [hep-ph]].


\bibitem{Pakhlova:2008zza}
  G.~Pakhlova {\it et al.}  [Belle Collaboration],
  Phys.\ Rev.\  D {\bf 77}, 011103 (2008)
  [arXiv:0708.0082 [hep-ex]].

\bibitem{Li:2007xr}
  G.~Li and Q.~Zhao,
  Phys.\ Lett.\  B {\bf 670}, 55 (2008)
  [arXiv:0709.4639 [hep-ph]].



\bibitem{bes-iii} D. M. Asner et al, ``Physics at BES-III¡±, Edited by K.T. Chao
and Y.F. Wang, Int. J. of Mod. Phys. {\bf A 24} Supplement 1, (2009)
[arXiv:0809.1869].

\bibitem{Feldmann:2000hs}
  T.~Feldmann and P.~Kroll,
  Phys.\ Rev.\  D {\bf 62}, 074006 (2000)
  [arXiv:hep-ph/0003096].


\end{thebibliography}
\end{document}